\documentclass[ twocolumn,aps,prd,   
               preprintnumbers,numbers,sort&compress,
               nofootinbib,
                            showpacs,
               colorlinks,
               linkcolor=blue,
               citecolor=blue]{revtex4-1}
   \newcommand{\exclude}[1]{}

\usepackage{graphicx,amsmath,amssymb,bm}
\usepackage{psfrag}
 \usepackage{feynmp}
\usepackage{hyperref}
\usepackage{enumitem}

\newcommand{\beq}{\begin{equation}}
\newcommand{\eeq}{\end{equation}}
\newcommand{\be}{\begin{eqnarray}}
\newcommand{\ee}{\end{eqnarray}}
   
\def\dd{ \,\mathrm{d} }

\def\+{\dagger}
 \def\la{\langle}
 \def\ra{\rangle}

\begin{document}

\title{ Dynamical Casimir Effect  in a small compact manifold   for the Maxwell vacuum.   }

\author{   Ariel R. Zhitnitsky}  

\affiliation{ Department of Physics \& Astronomy, University of British Columbia, Vancouver, B.C. V6T 1Z1, Canada} 

\begin{abstract}
We study  novel type of contributions to  the  partition function  of  the Maxwell system  defined on a small compact manifold ${\mathbb{M}}$ such as torus. These new terms   can not be described in terms of  the physical propagating photons with two transverse polarizations.  
Rather, these novel contributions  emerge as a result of   tunnelling events  when transitions occur between topologically different but physically  identical vacuum winding states. These new terms give an extra contribution to the Casimir pressure, yet to be measured.   

We argue that if the same  system   is considered in the background of a small external time-dependent magnetic field,  than there will be emission of photons from the vacuum, similar to the Dynamical Casimir Effect (DCE) when   real  particles are radiated  from the vacuum due to the   time-dependent  boundary conditions. The difference with conventional DCE is that the dynamics of the vacuum in our system is not related to the  fluctuations of the conventional degrees of freedom, the virtual photons. Rather, the radiation in our case  occurs as a result  of  tunnelling  events between topologically different but physically identical  $|k\ra$   sectors in a time -dependent background. We comment on relation of this novel effect with the well-known, experimentally observed, and theoretically understood  phenomena of the persistent currents in normal metal rings. We also comment on possible cosmological applications of this   effect.

\pacs{11.15.-q, 11.15.Kc, 11.15.Tk}
 
\end{abstract} 

\maketitle

\section{Introduction. Motivation.}\label{introduction}
 The main motivation for present studies is as follows.
 It has been recently argued \cite{Cao:2013na,Zhitnitsky:2013hba,Zhitnitsky:2014dra} that if  free Maxwell theory (without any interactions with charged particles) is   defined on a small compact manifold  than  some novel  terms in the partition function will emerge.  These terms are not related to the  propagating photons with two transverse physical polarizations, which are responsible for the conventional Casimir effect (CE)\cite{Casimir}. Rather, these novel terms occur as a result  of  tunnelling  events between topologically different but physically identical    $|k\ra$ topological sectors. These states play no role when the system is defined in infinitely large Minkowski space-time ${\mathbb{R}_{1,3}}$. But these states become important when   the system   is defined on a small  compact manifold. Without loosing any generality we shall call this manifold  ${\mathbb{M}}$, it could be the four-torus ${\mathbb{T}^4}$, 
 or it could be any other compact manifold with a  non-trivial mapping $\pi_1[U(1)]= \mathbb{Z}$. Precisely this non-trivial mapping for the Maxwell $U(1)$ gauge theory implies the presence of the topological sectors $|k\ra$ which play the key role in our discussions. The corresponding phenomenon  was coined as the topological Casimir Effect (TCE). \exclude{ A metallic cylinder (solenoid) covered by  caps  (and isolated from the solenoid) at small,  but finite temperature is a possible realization for such a manifold. We recover the conventional   Casimir pressure  (Casimir force between the caps at zero temperature) when the radius  of the cylinder becomes much larger than its width, in which case the topological tunnelling transitions  between $|k\ra$ sectors  are strongly suppressed, and can be ignored. }
 
 In particular,  it has been explicitly shown in  \cite{Cao:2013na}  that these novel terms in  the topological portion of the partition function ${\cal{Z}}_{\rm top}$  lead to a fundamentally new  contributions to the Casimir vacuum pressure, which can  not be expressed  in terms of conventional  propagating  physical degrees of freedom.   \exclude{  Instead,  the new vacuum contributions   appear  as a result  of tunnelling events between different topological sectors $|k\ra$. }
 Furthermore, the   ${\cal{Z}}_{\rm top}$  shows  many features 
 of   topologically ordered systems, which were  initially   introduced in context of condensed matter (CM) systems, see     recent reviews \cite{Cho:2010rk,Wen:2012hm,Sachdev:2012dq, Cortijo:2011aa, Volovik:2011kg}.  In particular, ${\cal{Z}}_{\rm top}$ demonstrates  the degeneracy of the system which can not be described in terms of any local operators  \cite{Zhitnitsky:2013hba}. Instead, such a degeneracy can be formulated in terms of  some non-local operators, see few comments on this classification in Appendix \ref{Appendix}. Furthermore, the infrared physics of the   system can be studied  in terms of   auxiliary topological non-propagating fields \cite{Zhitnitsky:2014dra} precisely in the same way as a topologically ordered system  can be  analyzed  in terms of  the Berry's connection (which is also emergent rather than a fundamental field).    Furthermore, the corresponding expectation value of the auxiliary topological filed  determines the phase of the system.
 
 As we review in section \ref{4d},  the relevant   vacuum fluctuations  which saturate the topological portion of the partition function ${\cal{Z}}_{\rm top}$      are formulated in terms of topologically non-trivial boundary conditions. 
 These configurations satisfy  the periodic boundary conditions on gauge  field up to a large gauge transformation
  such that the tunnelling transitions occur between physically identical but topologically distinct $|k\ra$ sectors.
 Precisely these field configurations generate   an  extra Casimir vacuum pressure in the system. What happens to this complicated vacuum structure when the system is placed into the background of external constant magnetic field  $B_{\rm ext}^z$? The answer on this question is known: the corresponding partition function ${\cal{Z}}_{\rm top}$ as well as all  observables, including the topological part of the Casimir pressure,  are highly sensitive to small magnetic field and demonstrate
   the $2\pi$ periodicity  with respect to   magnetic flux represented by parameter $\theta_{\rm eff}\equiv eSB_{\rm ext}^z$  where $S$ is the $xy$ area   of the system  ${\mathbb{T}}^4$.  This sensitivity to external magnetic field  is a result of  the quantum interference  of the external filed $ B_{\rm ext}^z$ with topological quantum fluctuations describing the tunnelling transitions between $|k\ra$ sectors. This strong ``quantum" sensitivity of the TCE should be contrasted  with conventional Casimir forces  which are  practically unaltered by the external    field due to very strong suppression $\sim B_{\rm ext}^2/m_e^4$, see   \cite{Cao:2013na} for the details.

The main goal of the present work is  to the study  the dynamics of these   vacuum  fluctuations 
 in the presence of a {\it  time-dependent } magnetic  field    $B_{\rm ext}^z(t)$.         We would like  to argue  in the present work that there  will be a   radiation of  real photons   emitted from  the   vacuum  (described  by the partition function ${\cal{Z}}_{\rm top}[B_{\rm ext}^z(t)]$)  as a  result   of this time dependent external source  $B_{\rm ext}^z(t)$. 

A simple intuitive picture of  this  emission can be explained as follows.  Imagine that we study conventional CE with static metallic plates. When these plates move or fluctuate, there will be emission from the vacuum, which is well known and well studied phenomenon   known as  the Dynamical Casimir Effect (DCE), see original papers \cite{DCE} and  reviews \cite{DCE-review}. The DCE has been observed  in recent experiments with superconducting circuits \cite{DCE-exp}. Basically, the virtual photons which are responsible for conventional Casimir pressure may become the real photons when the plate is moving or fluctuating. 

The novel effect which is the subject of this work is that   the topological configurations describing the tunnelling transitions between $|k\ra$ sectors  will be also modified  when there is a time dependent external influence on the system. This time dependent impact on the system can be realized by moving the plates of the original manifold  ${\mathbb{M}}$, in close analogy with DCE. The time-dependent impact may   also enter the system through the quantum interference of the external field with topological configurations saturating the partition function ${\cal{Z}}_{\rm top}$. Such quantum interference, as we mentioned above,  is practically absent in conventional   CE but is order of one  in the TCE. Therefore, this quantum interference gives us a unique chance to manipulate with the Maxwell vacuum  defined on  ${\mathbb{M}}$ using a time-dependent external  electromagnetic source. This effect      leads to the production of   the real photons with transverse polarizations  emitted from the topological quantum vacuum configurations saturating the partition function ${\cal{Z}}_{\rm top}$ in the presence of time-dependent magnetic field  $B_{\rm ext}^z(t)$.

As this  effect is very novel and quite  counter-intuitive, we would like to present one more additional  explanation supporting our claim that there will be emission of real photons from the Maxwell vacuum when the systems (described by the partition function ${\cal{Z}}_{\rm top}$)  is placed into the background of time-dependent source  $B_{\rm ext}^z(t)$. 

Our second explanation goes as follows.
The topological configurations which describe the tunnelling transitions are formulated in terms of  the  boundary conditions   on gauge field up to a large gauge transformations. These boundary conditions  correspond to  some persistent fluctuating currents which can flow along the  metallic   boundaries corresponding to the edges of ${\mathbb{M}}$. In fact, the possibility that such  persistent current may occur  in small   rings with topology ${\mathbb{S}^1}$ have been theoretically predicted long ago
\cite{persistent}, though with very different motivation from the one advocated in present work. Furthermore,  the corresponding  persistent non-dissipating currents in  different   materials have been experimentally observed in small rings ${\mathbb{S}^1}$, see reviews \cite{review_persistent}. Our comment here is that similar  currents flowing along the rings  
of ${\mathbb{S}^1}$  which represents the boundary  of ${\mathbb{M}}$ can be  interpreted  as a result of topological vacuum configurations intimately related to Aharonov Bohm phases when the system is defined on a topologically nontrivial manifold. 
We elaborate on this connection (between the persistent currents and our description in terms of the topological vacuum configurations) further in the text. 

The only comment we would like to make here  is as follows. The  persistent clockwise and anti-clockwise currents  cancel each other in case of vanishing  external magnetic field. This cancellation does not hold in the presence of a time-independent external magnetic field $B_{\rm ext}^z$ perpendicular to the ring ${\mathbb{S}^1}$, in which  case 
 the persistent current $I_0$ will be generated.   One can view this system as  generation of a static   magnetic moment 
${m}_{\rm ind}^z=I_0 S$. It is quite obvious now that if the external magnetic field $B_{\rm ext}^z(t) $ becomes a time-dependent function, the corresponding induced magnetic moment  ${m}_{\rm ind}^z(t)$  also becomes a   time-dependent function. The corresponding time-dependence in ${m}_{\rm ind}^z(t)$  obviously implies that  the system starts to radiate physical photons with typical angular distribution given by the magnetic dipole radiation. This radiation is ultimately related to the topological vacuum configurations describing the tunnelling transitions between   $|k\ra$ sectors. These vacuum configurations get modified in the presence of a time-dependent field $B_{\rm ext}^z(t) $, which is precisely the source for the radiation of physical photons.  
In all respects the idea is very similar to DCE with the ``only" difference is that the conventional virtual photons (responsible for the CE) do not interfere with external magnetic field $B_{\rm ext}^z(t) $, while the topological vacuum instanton-like configurations  (saturating the TCE) do. We coin the corresponding phenomenon of the emission of real photons from vacuum configurations   saturating ${\cal{Z}}_{\rm top}[B_{\rm ext}^z(t) ]$ in the presence of time dependent source $[B_{\rm ext}^z(t) ]$ the  non-static  (or dynamical)  topological Casimir effect (TCE) to discriminate it from the conventional  DCE.

  \exclude{\footnote{There has always hung a shadow over this question about vacuum fluctuations as there have always been suspicions that those vacuum fluctuations are not really  zero point fluctuations, but rather can be attributed to some other physics, see in particular the relatively recent paper  \cite{Jaffe:2005vp} where it has been argued that the conventional Casimir effect (CE) can be computed without even mentioning such a notion as the ``vacuum".  See, however,    Appendix  in 
\cite{Cao:2013na} with a number of arguments  suggesting  that the topological vacuum fluctuations, leading to the topological Casimir effect (TCE) which is the subject of the   present work are very real and very physical, and cannot be removed by any means such as subtraction or redefinition of observables.}}

  The structure of our presentation is as follows.   In next section \ref{4d} we review our previous results \cite{Cao:2013na,Zhitnitsky:2013hba,Zhitnitsky:2014dra} on construction of  the partition function ${\cal{Z}}_{\rm top}$
   describing the tunnelling transitions between $|k\ra$ sectors. We also explain how this partition function is modified in the presence of external static magnetic field $B_{\rm ext}^z $.    After that in section \ref{time} we generalize the construction to include the slow time-dependent fields, which allow us to compute the induced magnetic dipole moment of the system. This time dependent induced magnetic moment radiates real physical photons  from vacuum. We also elaborate   on relation of our construction with persistent currents in section \ref{interpretation}. Finally, in sections \ref{numerics}, \ref{radiation} we make few simple numerical,  order of magnitude estimates in order   to get some insights on  potential prospects of  measuring  the effect, which crucially depends on property of  degeneracy of the system. As this   feature of degeneracy   is crucial  for potential experimental studies of this effect, we make few comments  on this property   
   in Appendix \ref{Appendix}.  The corresponding description of the system (when it is characterized by a   global, rather than local  observables) is quite different from conventional classification scheme    when a  system is   characterized by  an expectation value of  a  local operator.  
   
   Our conclusion is   section \ref{conclusion} where we  speculate    on possible relevance   of this novel effect for cosmology when     appropriate topology is  $\pi_3[SU(3)]= \mathbb{Z}$ replacing the  nontrivial mapping $\pi_1[U(1)]= \mathbb{Z}$ considered in present work  for studying the Maxwell theory on a compact manifold.     To be more concrete, we speculate  that the de Sitter behaviour in inflationary epoch could be just inherent property of the topological sectors in QCD in expanding Universe, rather than a result of dynamics of some ad hoc  dynamical field such as inflaton. The emission of real physical degrees of freedom from the inflationary  vacuum in time dependent background (the so-called reheating epoch) in all respects is very similar to the effect considered in the present work when the real photons can be emitted from vacuum in the background of a time dependent magnetic field  $B_{\rm ext}^z(t)$.

  \section{Topological partition function}\label{4d}
 Our goal here is to review   the Maxwell system  defined  on a Euclidean 4-torus   with  sizes $L_1 \times L_2 \times L_3 \times \beta$ in the respective directions. It provides the infrared (IR) regularization of the system. 
 This IR regularization plays a key role in proper treatment of the  topological terms which are  related to tunnelling events 
 between topologically distinct but physically identical $|k\ra$ sectors.
  
 \subsection{Construction}\label{construction}
 We follow  \cite{Cao:2013na,Zhitnitsky:2013hba,Zhitnitsky:2014dra}  in our construction of the partition function ${\cal{Z}}_{\rm top}$ where it was employed    for  computation of  the corrections to the Casimir effect due to these novel type of topological fluctuations. The crucial point is that we impose the periodic boundary conditions on gauge $A^{\mu}$ field up to a large gauge transformation.
 In what follows we simplify our analysis by considering   a clear case with winding topological sectors $|k\ra$    in the z-direction only.  The classical instanton configuration in Euclidean space  which describes the corresponding tunnelling transitions can be represented as follows:
\be
\label{toppot4d}
A^{\mu}_{\rm top} = \left(0 ,~ -\frac{\pi k}{e L_{1} L_{2}} x_2 ,~ \frac{\pi k}{e L_{1} L_{2}} x_1 ,~ 0 \right),
\ee  
where $k$ is the winding number that labels the topological sector, and $L_{1}$, $L_{2}$ are the dimensions of the plates in the x and y-directions respectively, which are assumed to be much larger than the distance between the plates $L_3$. This terminology (``instanton") is adapted   from  similar 
 studies in 2d QED    \cite{Cao:2013na} where corresponding configuration in $A_0=0$ gauge describe the interpolation between pure gauge vacuum winding states $|k\ra$. We  use the same terminology and interpretation for 4d case  because (\ref{topB4d}) is the classical configuration saturating the partition function ${\cal{Z}}_{\rm top}$ in close analogy with 2d case as discussed  in  details in \cite{Cao:2013na}.This classical instanton-flux  configuration satisfies the periodic boundary conditions up to a large gauge transformation,  and provides a topological magnetic instanton-flux in the z-direction:
\be
\label{topB4d}
\vec{B}_{\rm top} &=& \vec{\nabla} \times \vec{A}_{\rm top} = \left(0 ,~ 0,~ \frac{2 \pi k}{e L_{1} L_{2}} \right),\\
\Phi&=&e\int dx_1dx_2  {B}_{\rm top}^z={2\pi}k. \nonumber
\ee
The Euclidean action of the system is quadratic and has the following  form  
\be
\label{action4d}
\frac{1}{2} \int \dd^4 x \left\{  \vec{E}^2 +  \left(\vec{B} + \vec{B}_{\rm top}\right)^2 \right\} ,
\ee
where $\vec{E}$ and $\vec{B}$ are the dynamical quantum fluctuations of the gauge field.  
We call the configuration given by  eq. (\ref{toppot4d}) the instanton-fluxes  describing the tunnelling events between topological sectors $|k\ra$. These configurations saturate the partition function (\ref{Z4d}) and should be interpreted as ``large" quantum fluctuations which change the winding states $|k\ra$, in contrast with ``small" quantum fluctuations which are topologically trivial and expressed in terms of conventional virtual photons saturating  the quantum portion of the partition function ${\cal{Z}}_{\rm quant}$.

The key point is that the  topological portion ${\cal{Z}}_{\rm top}$   decouples from quantum fluctuations,  ${\cal{Z}} = {\cal{Z}}_{\rm quant} \times {\cal{Z}}_{\rm top}$ such that the quantum fluctuations do not depend on topological sector $k$ and can be computed in topologically trivial sector $k=0$.
Indeed,  the cross term 
\be
\int \dd^4 x~ \vec{B} \cdot \vec{B}_{\rm top} = \frac{2 \pi k}{e L_{1} L_{2}} \int \dd^4 x~ B_{z} = 0 
\label{decouple}
\ee
vanishes  because the magnetic portion of quantum fluctuations in the $z$-direction, represented by $B_{z} = \partial_{x} A_{y}  - \partial_{y} A_{x} $, is a periodic function as   $\vec{A} $ is periodic over the domain of integration. 
This technical remark in fact greatly simplifies our  analysis as the contribution of the physical propagating photons 
is not sensitive to the topological sectors $k$. This is,  of course,  a specific feature  of quadratic action 
 (\ref{action4d}), in contrast with non-abelian  and non-linear gauge field theories where quantum fluctuations of course depend on topological $k$ sectors.  
  
The classical action  for configuration (\ref{topB4d}) takes the form 
\be
\label{action4d2}
\frac{1}{2}\int \dd^4 x \vec{B}_{\rm top}^2= \frac{2\pi^2 k^2 \beta L_3}{e^2 L_1 L_2}
\ee
To simplify our analysis further in  computing  ${\cal{Z}}_{\rm top}$ we consider a geometry where $L_1, L_2 \gg L_3 , \beta$ similar to construction relevant for the Casimir effect.   
 In this case our system   is closely related to 2d Maxwell theory by dimensional reduction: taking a slice of the 4d system in the $xy$-plane will yield precisely the topological features of the 2d torus considered in  great details in  \cite{Cao:2013na}.
  Furthermore, with this geometry our simplification (\ref{topB4d}) when we consider exclusively the magnetic instanton- fluxes in $z$ direction is justified as the corresponding classical action (\ref{action4d2}) assumes a minimal  possible values.  With this assumption we can consider very small temperature, but still we can not take a formal limit $\beta\rightarrow\infty$  in our final expressions
 as a result of our technical constraints in the system. 
      
With these additional simplifications   the topological partition function becomes  \cite{Cao:2013na,Zhitnitsky:2013hba,Zhitnitsky:2014dra} :
\be
\label{Z4d}
{\cal{Z}}_{\rm top} = \sqrt{\frac{2\pi \beta L_3}{e^2 L_1 L_2}} \sum_{k\in \mathbb{Z}} e^{-\frac{2\pi^2 k^2 \beta L_3}{e^2 L_1 L_2} }= \sqrt{\pi \tau} \sum_{k\in \mathbb{Z}}e^{-\pi^2 \tau k^2}, ~~~~
\ee
where we introduced the dimensionless parameter
\be
\label{tau}
\tau \equiv {2 \beta L_3}/{e^2 L_1 L_2}.
\ee
Formula (\ref{Z4d})   is essentially the dimensionally reduced expression for  the topological partition function  for 2d 
 Maxwell theory analyzed in \cite{Cao:2013na}. 
   One should note that the normalization factor $\sqrt{\pi \tau}$ which appears in eq. (\ref{Z4d}) does not depend on topological sector $k$, and essentially it represents our convention of the  normalization   ${\cal{Z}}_{\rm top}
\rightarrow 1$ in the limit $L_1L_2\rightarrow \infty$ which corresponds to  a convenient set up for the  Casimir -type experiments as discussed in  \cite{Cao:2013na,Zhitnitsky:2013hba,Zhitnitsky:2014dra}. The simplest way to  demonstrate that ${\cal{Z}}_{\rm top} \rightarrow 1$ in the limit $\tau\rightarrow 0$ is to use the  dual representation  (\ref{Z_dual1}), see below.

\subsection{External static magnetic field  }\label{magnetic}

In this section we want to generalize our results   for the Euclidean Maxwell system in the presence of the external magnetic field. Normally, in the conventional quantization of electromagnetic fields in infinite Minkowski space, there is no \emph{direct} coupling    between fluctuating vacuum photons and an external magnetic field as a consequence of linearity of the Maxwell system. The coupling with  fermions   generates  a negligible effect $\sim \alpha^2B_{ext}^2/m_e^4$ as the non-linear Euler-Heisenberg Effective Lagrangian  suggests, see \cite{Cao:2013na} for the details and numerical estimates.  In contrast with conventional   photons, the the external magnetic field does couple with topological fluctuations (\ref{topB4d}).  It  leads to the effects of order of unity as a result of interference of the external magnetic field with topological fluxes $k$.

The corresponding partition function can be easily constructed for external magnetic field $B_{z}^{\rm ext}$ pointing along $z$ direction, as 
the crucial technical element on decoupling of the background fields  from quantum fluctuations assumes the same form (\ref{decouple}).  In other words, the physical propagating photons with non-vanishing momenta are not sensitive to the topological $k$ sectors, nor to the external uniform magnetic field, similar to our discussions after eq.(\ref{decouple}).

The classical action for configuration in the presence of the uniform static external magnetic field $B_{z}^{\rm ext}$ therefore takes the form 
\be
\label{B_ext}
\frac{1}{2}\int \dd^4 x  \left(\vec{B}_{\rm ext} + \vec{B}_{\rm top}\right)^2=  \pi^2\tau\left(k+\frac{\theta_{\rm eff}}{2\pi} \right)^2
\ee
where $\tau$ is defined by(\ref{tau})   and  the effective theta parameter $\theta_{\rm eff} \equiv e L_1L_2 B^z_{\rm ext}$ is expressed in terms of the original external magnetic field $B^z_{\rm ext}$.
Therefore, the partition function in the presence of the uniform magnetic field can be easily reconstructed from (\ref{Z4d}),   and it is given by  \cite{Cao:2013na,Zhitnitsky:2013hba,Zhitnitsky:2014dra} 
\be 
\label{Z_eff}
  {\cal{Z}}_{\rm top}(\tau, \theta_{\rm eff})
 =\sqrt{\pi\tau} \sum_{k \in \mathbb{Z}} \exp\left[-\pi^2\tau \left(k+\frac{\theta_{\rm eff}}{2\pi}\right)^2\right].~~
\ee
This system in what follows will be referred as the topological vacuum (  $\cal{TV}$) because the  propagating degrees of freedom, the photons with two transverse polarizations,   
completely decouple from  ${\cal{Z}}_{\rm top}(\tau, \theta_{\rm eff})$. 

  The dual representation for the partition function is obtained by applying the Poisson summation formula  such that (\ref{Z_eff}) becomes 
  \be 
\label{Z_dual1}
  {\cal{Z}}_{\rm top}(\tau, \theta_{\rm eff})
  = \sum_{n\in \mathbb{Z}} \exp\left[-\frac{n^2}{\tau}+in\cdot\theta_{\rm eff}\right]. 
  \ee
 Formula (\ref{Z_dual1})  justifies our notation for  the effective theta parameter $\theta_{\rm eff}$ as it enters the partition function in combination with integer number $n$. One should emphasize that integer  number $n$ in the dual representation (\ref{Z_dual1}) is not the integer magnetic flux $k$ defined by eq. 
(\ref{topB4d}) which enters the original partition function (\ref{Z4d}). Furthermore,  the $\theta_{\rm eff}$ parameter which enters (\ref{Z_eff}, \ref{Z_dual1}) is not a fundamental $\theta$ parameter which is normally introduced into the Lagrangian  in front of  $\vec{E}\cdot\vec{B}$ operator. Rather, this parameter  $\theta_{\rm eff}$ should be understood as an effective parameter representing the construction of the  $|\theta_{\rm eff}\ra$ state for each slice with non-trivial $\pi_1[U(1)]$ in four dimensional system. In fact, there are three such  $\theta_{\rm eff}^{M_i}$  parameters representing different slices and corresponding external magnetic fluxes. There are similar three $\theta_{\rm eff}^{E_i}$
parameters representing the external electric fluxes   as discussed  in \cite{Zhitnitsky:2013hba}, such that total number of $\theta$ parameters classifying the system equals six, in agreement with total number of hyperplanes in four dimensions. 
We shall not elaborate on this classification in the present work. In this work we limit ourselves with a single $\theta_{\rm eff}$ parameter entering (\ref{Z_eff}), (\ref{Z_dual1}), and corresponding to the magnetic external field $ B^z_{\rm ext} $ pointing  in $z$ direction.

\section{Induced magnetic dipole moment and  $E\&M$ radiation}\label{time}
The main goal of this section  is to estimate the  induced magnetic dipole moment of the system in a time dependent background. 
First, in section   \ref{basics} we derive  a   formula for induced magnetic moment in the background  of a uniform static external magnetic field.  We shall discuss  a different geometry in this section (in comparison with 4-torus discussed in previous review section \ref{4d}). This is because  our generalization for  a time dependent background  cannot  be  consistently introduced on 4 torus.  As the corresponding formula  for the induced magnetic moment  plays an important role in  the present work, we  offer a complementary interpretation of the same expression   in section \ref{interpretation}  in terms of the dynamics on the boundaries of the system, rather than in terms of the bulk-instantons (\ref{toppot4d}), (\ref{topB4d}).    In section  \ref{numerics} we provide   some numerical estimates, and make few comments on relation with experimentally observed persistent currents.   Finally, we  generalize the expression for induced magnetic moment for a slowly varying field. 
A   time dependent magnitude for the obtained magnetic moment  automatically implies the radiation of the real physical photons from this system as we show in sections \ref{radiation}.

\subsection{Magnetization of the system. The basics.}\label{basics}
Our goal here is to construct the topological portion  ${\cal{Z}}_{\rm top} $  for the partition function similar to (\ref{Z_eff}), (\ref{Z_dual1}), but for a different geometry. To be more specific, we want to consider a solenoid (cylinder) with opened ends   to have an option to place our system under influence of time variable magnetic field (which is not possible for 4-torus). This geometry also  gives  us an opportunity to    discuss the relation between the topological currents derived from ${\cal{Z}}_{\rm top} $ in next section \ref{interpretation} and persistent currents considered long ago \cite{persistent}. To proceed with this goal
 we consider a cylinder  with cross section of area  $\pi R^2$  which replaces  the  area $L_1L_2$ from original computations \cite{Cao:2013na} on 4 torus. Furthermore, for this geometry it is  convenient to consider the instanton solution describing the transitions between topological $|k\ra$ sectors in Cylindrical coordinates rather than in Cartesian  coordinates (\ref{toppot4d}). The corresponding ``instanton"-like   configuration   describes the the same physics as we discussed before in section \ref{4d}, and it  is given by 
\be
\label{cylindrical}
\vec{A}_{\rm top}=\frac{kr}{eR^2}\hat{\phi}, ~~r<R; ~~~~~~~~~ \vec{A}_{\rm top}=\frac{k}{er}\hat{\phi}, ~~r\geq R~~~
\ee
where $\hat{\phi}$ is unit vector in $\hat{\phi}$ direction  in Cylindrical coordinates. 
By construction, the vector potential  $\vec{A}_{\rm top}=\frac{k}{e}\vec{\nabla}\phi$ is   a pure gauge field on the boundary with nontrivial winding number $k$.  The topological magnetic flux in the $z$ direction is defined similar to (\ref{topB4d}),
and it is given by
\be
\label{replace1}
&&\vec{B}_{\rm top} =     \frac{2  k}{e R^2} \hat{z} ,~~r<R;  ~~~~~~ \vec{B}_{\rm top} =0, ~~r>R, \\
 &&\Phi=e\int dx_1dx_2  {B}_{\rm top}^z= e\oint_{r=R}  \vec{A}_{\rm top} \cdot d\vec{l}={2\pi}k. \nonumber
\ee
The corresponding formulae for   ${\cal{Z}}_{\rm top}(\tau, \theta_{\rm eff})$ replacing (\ref{Z_eff}), (\ref{Z_dual1}) with new geometry assume  the form:
\be
\label{replace}
&& {\cal{Z}}_{\rm top}(\tau, \theta_{\rm eff})
 =\sqrt{\pi\tau} \sum_{k \in \mathbb{Z}} \exp\left[-\pi^2\tau \left(k+\frac{\theta_{\rm eff}}{2\pi}\right)^2\right]\nonumber \\
   &&= \sum_{n\in \mathbb{Z}} \exp\left[-\frac{n^2}{\tau}+in\cdot\theta_{\rm eff}\right], ~~~~
 \tau \equiv {2 \beta L_3}/{e^2 \pi R^2},  ~~
 \ee
where $L_3$ is the length of the cylinder,  $\beta$ is the inverse temperature, and effective theta parameter $\theta_{\rm eff} \equiv e \pi R^2 B^z_{\rm ext}$ is expressed in terms of the original external magnetic field $B^z_{\rm ext}$ similar to (\ref{B_ext}).

Few comments are in order. First, the topological portion of the partition function for the cylinder given  by (\ref{replace}) plays the same role 
  as (\ref{Z_eff}), (\ref{Z_dual1}) plays for the 4-torus. There are few differences, though. In case for the 4-torus we could, in principle, introduce     6 different $\theta_i$ parameters corresponding to 6 different hyperplanes and 6 different nontrivial mappings $\pi_1[U(1)]= \mathbb{Z} $ for each slice  in 4d Euclidean space. In contrast, with new geometry   there is  just one  type of magnetic fluxes  (\ref{replace1}).    Another difference is that  our treatment of 4-torus in previous section   neglects all other types of instantons. This was achieved  by imposing geometrical condition $L_1,L_2\gg \beta, L_3$ to  guarantee  that action (\ref{action4d2})  assumes a minimal possible value (maximal contribution to the partition function), while other types of instantons would produce  parametrically smaller contribution to the partition function  ${\cal{Z}}_{\rm top}(\tau, \theta_{\rm eff})$. In present case   
  we do not have any other types of instantons which may contribute  to ${\cal{Z}}_{\rm top}(\tau, \theta_{\rm eff})$. However, we must assume that $L_3\gg R$ to justify our approximate for (finite $L_3$) solution  (\ref{cylindrical}), (\ref{replace1}) when  a contribution from   outside  region  can be neglected in the action (\ref{replace}). 
 Finally, as we already mentioned, the normalization factor $\sqrt{\pi\tau}$  in ${\cal{Z}}_{\rm top}(\tau, \theta_{\rm eff})$
does not depend on topological sectors $k$, and in fact, is a matter of convention.  It is more appropriate for the given geometry to define  ${\cal{Z}}'_{\rm top}  ={\cal{Z}}_{\rm top}/ \sqrt{\pi\tau} $ such that ${\cal{Z}}'_{\rm top}(\tau\rightarrow\infty, \theta_{\rm eff}=0) \rightarrow 1$. However, we opted to preserve our original normalization because our results which follow do not depend on this  convention. 
\exclude{
In normal circumstances the topological portion of the partition function $ {\cal{Z}}_{\rm top}\rightarrow 1 $ and  plays no role in  physics\footnote{In particular, $ {\cal{Z}}_{\rm top}$ gives   parametrically small contributions into thermodynamical observables, such as entropy \cite{Zhitnitsky:2013hba}, where it may produce subleasing (but universal) constant contribution $S=\ln 2$, in contrast with conventional  expression for $S\sim T^3V$  generated by  $ {\cal{Z}}_{\rm quant}$ and proportional to the volume of the system.  It also gives parametrically small contribution to the Casimir pressure when one uses  typical  for CE parameters  with $L_3/R\leq 10^{-4}$, see  \cite{Cao:2013na}.}.  The  goal of the present work is to argue that the new and interesting physics hidden  in $ {\cal{Z}}_{\rm top}$  could  be studied by exciting the topological vacuum configurations using a  quantum interference with  external time-dependent field $B_{z}^{\rm ext}(t)$.  This is the subject of next section.
}

Our next step  is to analyze 
the magnetic response of the system under influence of the external  magnetic field.  The idea behind these studies is the observation that the external magnetic field acts as an effective $\theta_{eff}$ parameter as eq. (\ref{replace}) suggests.  
Therefore, one can differentiate with respect to this parameter to compute  
 the induced magnetic field  
\be 
\label{B_ind}
&\,& \langle B_{\rm ind} \rangle = -\frac 1 {\beta V}\frac{\partial \ln \mathcal{Z}_{\rm top}}{\partial B_{\rm ext}^z}=
-\frac{e}{\beta L_3}\frac{\partial \ln\mathcal{Z}_{\rm top}}{\partial\theta_{\rm eff}}\\
&=& \frac{\sqrt{\tau\pi}}{\mathcal{Z}_{\rm top}}\sum_{k\in\mathbb{Z}}\left(B_{\rm ext}+\frac{2 k}{ R^2e}\right)\exp{\left[-\tau\pi^2(k+\frac{\theta_{\rm eff}}{2\pi})^2\right]}.\nonumber
\ee

As one can see from (\ref{B_ind}),  our definition of the induced field  accounts for the total field which includes both terms: the external part as well as the topological portion of the field. In the absence of the external field ($B^{\rm ext}=0$), the series is antisymmetric under $k\rightarrow -k$ and $\langle B_{\rm ind} \rangle$ vanishes. It is similar to the vanishing expectation value of the topological density  in gauge theories  when $\theta=0$. One could anticipate this result from symmetry arguments as the theory must respect $\cal{P}$ and $\cal{CP}$ invariance at $\theta=0$.

The expectation value of the induced magnetic field exhibits  the $2\pi$ periodicity from the partition function and it reduces to triviality whenever the amount of skewing results in an antisymmetric summation, i.e. $\langle B_{\rm ind}\rangle = 0$ for $\theta_{\rm eff}\in\{2n\pi:n\in\mathbb{Z}\}$. The point $\theta_{\rm eff} =\pi$ deserves special attention as this point corresponds to the degeneracy, see  \cite{Zhitnitsky:2013hba} with detail discussions. This degeneracy can not be detected by an  expectation value of any local operator, but rather is classified by a nonlocal operator,   similar to studies of the topological insulators at $\theta=\pi$.

\begin{figure}[htb]
  \center{\includegraphics[width = 0.50\textwidth]{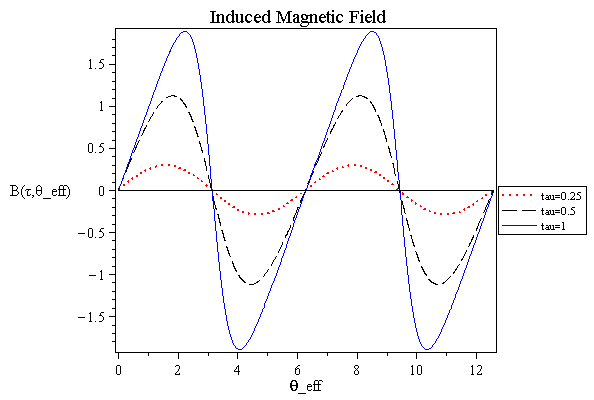}}
  \caption{\label{fig:bfield} A numerical plot  of the induced magnetic field in units  $\frac{c}{\pi R^2 e}$ as a function of external flux $\theta_{eff}$.  The same plot represents the induced magnetic moment $-\langle m^z_{\rm ind} \rangle$ in units $\frac{e L_3 c}{4\pi \alpha}$, see text with details. Unit magnetic flux corresponds to $\theta_{eff}=2\pi$. The plot is adapted from  \cite{Cao:2013na}. }
\end{figure}
We now return to analysis of eq. (\ref{B_ind}). 
The topological  effects, as expected,  are exponentially suppressed at $\tau\ll 1$ and $\tau\gg 1$   according to eq.(\ref{replace}). The effect  is much   more pronounced in the range where $\tau\simeq 1$, see Figure \ref{fig:bfield}, where we plot  the induced magnetic field in units  $(\pi R^2e)^{-1}$ as a function of external flux $\theta_{\rm eff}$ (point $\theta_{\rm eff}=\pi$ should be considered separately as we mentioned above).   

Important comment  here is that the induced magnetic field defined as (\ref{B_ind}) can be thought of as the magnetization of the system per unit volume, i.e. $\la M\ra=-\la B_{\rm ind}\ra$, as the definition for 
$\la M\ra$ is identical to (\ref{B_ind}) up to a minus sign because it enters the Hamiltonian as $H=-\vec{m}_{\rm ind}\cdot \vec{B}_{\rm ext}$. Therefore, we arrive to the following expression for the induce magnetic moment of the system  in the presence of 
external magnetic field $B_{\rm ext}^z$, 
\be 
\label{m_ind}
&\,& \langle m^z_{\rm ind} \rangle = \frac 1 {\beta }\frac{\partial \ln \mathcal{Z}_{\rm top}}{\partial B^{\rm ext}_z}=
 -\langle B_{\rm ind} \ra L_3 \pi R^2 \\
&=&- \frac{L_3 2\pi}{e}\frac{\sqrt{\tau\pi}}{\mathcal{Z}_{\rm top}}\sum_{k\in\mathbb{Z}}\left(\frac{\theta_{\rm eff}}{2\pi}+ k\right)\exp{\left[-\tau\pi^2(k+\frac{\theta_{\rm eff}}{2\pi})^2\right]}.\nonumber
\ee
One can view Fig. \ref{fig:bfield} as a plot for the induced magnetic moment in units $\frac{e L_3 c}{4\pi \alpha}$ which represents correct dimensionality $\frac{\rm e\cdot  cm^2}{\rm s}$. 

   \subsection{Interpretation}\label{interpretation}
   
   As formula (\ref{m_ind}) plays a key role in our discussions,  we would like to interpret the same expression for $\langle m^z_{\rm ind} \rangle$ but in different terms. To be more precise: equation (\ref{m_ind}) describes the magnetization properties of the system in terms of the vacuum configurations  describing the tunnelling transitions   between topological sectors $|k\ra$.   The corresponding partition function ${\cal{Z}}_{\rm top}(\tau, \theta_{\rm eff})$ which governs these vacuum processes is given by (\ref{replace}). 
   A non-trivial behaviour of the magnetization of the system in terms of the induced magnetic dipole moment (\ref{m_ind})   is direct  consequence of the basic properties of the partition function ${\cal{Z}}_{\rm top}(\tau, \theta_{\rm eff})$. We would like to understand the same properties in more intuitive way in terms of the fluctuating currents which unavoidably  will be generated on the boundaries, as we discuss below.
   
   Indeed, the cross term in the effective action (\ref{B_ext}) which describes coupling of the external field with topological instanton-like  configuration can be represented as follows
    \be
\label{B_cross}
 \int \dd^4 x  \left(\vec{B}_{\rm ext} \cdot \vec{B}_{\rm top}\right)=   \int \dd^4 x ~\vec{A}_{\rm ext}\cdot \left(\vec{\nabla}\times \vec{B}_{\rm top}\right),~~
\ee
  where we neglected a  total  divergence term.  The cross term written in the form (\ref{B_cross}) strongly suggests that 
   $\left(\vec{\nabla}\times \vec{B}_{\rm top}\right)$ can be interpreted as a steady current flow along the boundary.
   Indeed, 
    \be
\label{j}
       \vec{j}_{\rm top}(k) = -\vec{\nabla}\times \vec{B}_{\rm top} =-\delta (r-R)\frac{2k}{eR^2}\hat{\phi},
    \ee 
   where we used expression (\ref{replace}) for $ \vec{B}_{\rm top}$ describing the tunnelling transition to $|k\ra$ sector. Formula (\ref{j}) is very   suggestive and implies  that the vacuum transitions formulated in terms of the fluxes-instantons (\ref{cylindrical}), (\ref{replace1}) can be also interpreted in terms of accompanied fluctuating  topological currents (\ref{j}). The total current  (in topological $k$ sector)  which flows along the infinitely thin  boundary  of a  cylinder radius $R$ and length $L_3$ in  our ideal system is given by
     \be
\label{J}
    {J}^{\phi} (k) = \int_0^{L_3} dz\int dr  {j}^{\phi}_{\rm top}    =-\frac{2kL_3}{eR^2}.
           \ee 
This current  in topological $k$ sector  produces the following contribution to the magnetic dipole moment:
\be
\label{m_ind1}
m_{\rm ind}^z (k)= \pi R^2 J^{\phi}(k)= -\frac{2\pi  L_3}{e}k.
\ee
This formula   precisely reproduces the term   proportional to $k$ in the parentheses in eq. (\ref{m_ind}) which was originally derived  quite  differently, see previous subsection \ref{basics}. 

 Few comments on eq. (\ref{J}). 
The expectation value $\la J^{\phi} \ra $ of the  topological current obviously vanishes at zero external magnetic field when one sums over all topological $k$-sectors, in agreement with our previous expression (\ref{m_ind}) with $\theta_{\rm eff}=0$. 
  It is quite obvious that  the topological  currents have   pure quantum   nature as they  effectively represent the instantons  describing the tunnelling transitions in the path integral computations. The currents   could have  clockwise or anticlockwise direction,  depending on sign of integer number $k$, similar to fluctuating  instanton solutions  (\ref{cylindrical}),  saturating the topological portion of the partition function (\ref{replace}). 
  
   Furthermore, one can explicitly check that the cross term (\ref{B_cross}) computed   in terms of the boundary current $  \vec{j}_{\rm top}(k) $ exactly reproduces the corresponding term in the action for the partition function (\ref{m_ind}) computed in terms of the bulk instantons (\ref{cylindrical}), (\ref{replace}). Indeed, 
      \be
\label{B_cross1}
   \int&& d^4 x\vec{A}_{\rm ext}\cdot \left(\vec{\nabla}\times \vec{B}_{\rm top}\right)=\frac{2k (2\pi\beta L_3)}{eR^2}\int rdr \delta (r-R)A^{\phi}_{\rm ext}\nonumber\\
 &&=\frac{2k (2\pi\beta L_3)}{eR^2} \cdot   \left(\frac{B^z_{\rm ext}R^2}{2}\right)   =2\tau \pi^2 k\left(\frac{\theta_{\rm eff}}{2\pi}\right),
      \ee
   where the  vector potential   ${A}^{\phi}_{\rm ext}(r)= \frac{r}{2}B_{\rm ext}^z $  corresponds  to the external uniform magnetic field in cylindrical coordinates. Our  final result in eq. (\ref{B_cross1}) is  expressed in  terms of the external flux $\theta_{\rm eff}\equiv e\pi R^2B_{\rm ext}^z$ and dimensionless parameter $\tau \equiv {2 \beta L_3}/{e^2 \pi R^2}$. One can explicitly see that the  cross term in action in eq. (\ref{m_ind}) is reproduced by eq. (\ref{B_cross1}) derived in terms of the boundary currents, rather then in terms of the bulk instantons.  
   
   The classical instanton action is also reproduced in terms of the boundary currents. Indeed, by substituting the expression for the current (\ref{j}) to the classical action one arrives to 
      \be
\label{B_current}
&&\frac{1}{2}  \int \dd^4 x  \left(\vec{B}_{\rm top} \cdot \vec{B}_{\rm top}\right)=  \frac{1}{2} \int \dd^4 x ~\vec{A}_{\rm top}\cdot \left(\vec{\nabla}\times \vec{B}_{\rm top}\right),\nonumber \\
&=&\frac{k (2\pi\beta L_3)}{eR^2} \int rdr \delta (r-R)A^{\phi}_{\rm top}(r)=\tau \pi^2 k^2, 
\ee
which is precisely the expression for classical  instanton   action entering the topological portion of the  partition function (\ref{m_ind}).

The basic point of our discussions in this section is that the expression for the induced magnetic moment (\ref{m_ind}) can be understood  in terms of the topological currents flowing along the boundaries of the system. However, 
 the origin  of the phenomena is not these currents but  the presence of the 
  topological $|k\ra$ sectors in Maxwell $U(1)$ electrodynamics when it is formulated on a compact manifold with nontrivial  
  mapping $\pi_1 [U(1)]=\mathbb{Z}$.  Such $|k\ra$ sectors exist and transitions between them always occur  even if  charged   particles are not present in the system.  The coupling of the non-trivial gauge configurations describing the transitions between the $|k\ra$ sectors  with charged particles in the presence of external magnetic field leads to such pronounced effect  as persistent currents \cite{persistent,review_persistent}.   The secondary, rather than fundamental
  role of the charged matter particles  in this phenomenon manifests, in particular, in a fact that $\mathcal{Z}_{\rm top}$ generates an extra  contribution to the Casimir vacuum pressure even at zero  external magnetic field. At the same time the persistent current can not be  generated at $B_{\rm ext}=0$ as clockwise and counterclockwise currents cancel each other. 
  
  Furthermore, the persistent currents in the original works  \cite{persistent,review_persistent} were introduced as a response 
  of the electrons (residing on the  ring) on external magnetic flux with nontrivial Aharonov -Bohm phase. In contrast,  the non-trivial gauge configurations (and accompanied topological currents (\ref{j}) in $k$ sectors)  in our  system are generated even when no external field nor  corresponding external Aharonov Bohm vector potential are present in the system. In addition, the correlation length in conventional persistent currents \cite{persistent,review_persistent} is determined by dynamics of the electrons residing on the  ring, while in our case it is determined by the dynamics of the vacuum  described by the  partition function ${\mathcal{Z}_{\rm top}}$.  The corresponding topological fluctuations (described in terms of  the instantons (\ref{replace})) also generate the  persistent topological currents on the boundary (\ref{j}). However, this additional contribution should be treated separately from conventional persistent currents  \cite{persistent,review_persistent}, as it is absolutely independent contribution which is generated due to the unavoidable coupling of topological gauge  configurations  with charged particles on the boundary of the system\footnote{In many respects  this situation is very similar to  QCD when the presence of the topological sectors is absolutely  fundamental basic property of the gauge system. At the same time,  very pronounced  consequences of this  fundamental feature are expressed in terms of the matter fermi fields, rather than in terms of original gauge configurations. These well noticeable   properties of the system   are basically the consequence  of  the Index Theorem which states that the  fermions in the background  of nontrivial gauge configurations have chiral zero modes.  Precisely these zero modes  play extremely important role in explanation of many effects such as generation of the chiral condensate in QCD, the  resolution of the so-called the $U(1)_A$ problem, etc. However, the root, the origin of these properties of the system is the presence of topologically non-trivial gauge configurations, while the matter fields play the secondary role.}. 
  
  Finally, the induced magnetic moment (\ref{m_ind1}) due to the topological currents flowing on the boundary is quantized.
  Indeed, $m_{\rm ind}^z/L_3$  assumes only integer numbers in units of $\frac{2\pi}{e}$. This is because the corresponding induced currents always accompany the quantized instanton -like fluctuations (\ref{replace1}). In contrast, a similar induced magnetic moment due to the conventional persistent currents  \cite{persistent,review_persistent} is not quantized, and can assume any value.
  
  To conclude this subsection we would like to comment that it is quite typical in condensed matter physics that the topologically ordered systems exhibit such a complementary formulation in terms of the physics on the boundary.  Our system ($\cal{TV}$) can be  also   thought as a topologically ordered system    as argued in \cite{Zhitnitsky:2013hba,Zhitnitsky:2014dra} because  it demonstrates  a number of specific features which are inherent properties of topologically ordered systems.  In particular, $\mathcal{Z}_{\rm top} (\tau, \theta_{\rm eff})$ demonstrates the degeneracy of the system which can not be described in terms of any local operators. Furthermore, the infrared physics of the system can be studied in terms of auxiliary topological non-propagating fields precisely in the same way as a topologically ordered system 
  in condensed matter physics can be analyzed in terms of the Berry's connection.  Therefore, it is not a surprise that we can reformulate the original instanton fluctuations saturating $\mathcal{Z}_{\rm top} (\tau, \theta_{\rm eff})$ in terms of  the boundary persistent currents which always accompany these instanton transitions.

\subsection{Numerical estimates}\label{numerics}
We want to make some simple numerical estimates  by comparing the magnetic induced moment $ \langle {m}^z_{\rm ind} \rangle$  from eq. (\ref{m_ind}) with corresponding expression $m_{\rm persistent}\simeq \pi I_0R^2$ with measured persistent current $I_0$. One should emphasize that the conventional persistent currents are highly sensitive to the properties  of the material.  More than that, 
the  properties of the condensed matter samples  essentially determine the magnitude of the measured currents. At the same time, in all our discussions above we assume that the ``ideal" boundary conditions can be  arranged, such that the Maxwell vacuum defined on a compact manifold is well described by the partition function (\ref{Z_eff}), (\ref{Z_dual1}).   Moreover, as we discussed in previous section \ref{interpretation} the topological boundary currents (\ref{j}) in our framework should be considered as an independent additional contribution to conventional persistent currents. Therefore, the corresponding numerical estimates taken from early work  \cite{persistent-exp}   are presented here for  demonstration purposes only.

The measurement of the typical persistent current was reported in \cite{persistent-exp}. 
The measurements were performed on single gold rings with diameter  $2 R=$2.4 and 4 $\mu$m
at a base temperature of 4.5 mK. Reported values for the currents are $I_0= $ 3 and 30 nA for these two   rings.    It should be contrasted with expected current $\sim 0.1$  nA. We are not in position to comment on this discrepancy, as the effect is basically determined by the properties of the material, which is not subject of the present work.  
Our goal is in fact quite different. We want to compare the magnetic moment which is induced due to this persistent current with 
induced magnetic moment due to the topological vacuum configurations $ \langle {m}^z_{\rm ind} \rangle$  from eq. (\ref{m_ind}). Numerically, a   magnetic moment for the largest observed current (30 nA) 
can be estimated as follows
\be
\label{observed}
m_{\rm persistent}&\simeq& \pi I_0R^2\sim \pi (30~ {\rm nA})\cdot \left(\frac{2.4~ \mu {\rm m}}{2}\right)^2 \nonumber \\
&\sim& 0.7 \cdot10^4 \left(\frac{e ~ {\rm cm}^2}{{\rm s}}\right).
\ee
It is instructive to compare this moment  with fundamental Bohr magneton   $\mu_B=e\hbar/2m_e\simeq 0.6 \cdot  \left(\frac{e ~ {\rm cm}^2}{{\rm s}}\right)$, which provides a crude  estimate of a number of effective degrees of freedom $\sim 10^4$ which generate  the persistent current $I_0$ for this specific sample. This estimate should be taken with some precaution because the effect of persistent currents is entirely determined by the properties of the material (such as the electron phase coherence length $l_{\phi}$) which is beyond of the scope of the present work.   

Before we estimate $ \langle {m}^z_{\rm ind} \rangle$  from eq. (\ref{m_ind}) to compare it with (\ref{observed}) we would like to get some insights  about the  numerical magnitude of the  dimensionless parameter $\tau$  for the   ring with parameters used in the estimate (\ref{observed}),
\be
\label{tau-exp}
\tau \equiv \frac{2 \beta L_3}{e^2 \pi R^2}\sim \frac{2 (0.1 \mu {\rm m}) (0.6 {\rm cm}) }{4\pi^2\alpha}\left(\frac{2}{2.4~ \mu {\rm m}}\right)^2 \gg 1,~~~
\ee
where we use for $L_3\sim 0.1 \mu {\rm m} $ and $\beta\sim 0.6$ cm which corresponds to the temperature $T\simeq 300 $ mK
below which  $l_{\phi}$ is sufficiently large and temperature independent\footnote{\label{AB}One should remark here that there are related effects  when  the entire system  can   maintain   the  Aharonov Bohm phase coherence  at    very  high temperature $T\simeq  79 $ K \cite{persistent-temp}.}. 
Large magnitude of $\tau$ implies that for the chosen parameters for the system the vacuum transitions between the topological sectors are strongly suppressed   as the expression for the partition function  (\ref{Z_eff}), (\ref{Z_dual1}) states.  In this regime the effect (\ref{observed}) is entirely determined by conventional mechanism \cite{persistent}, \cite{review_persistent}.

If somehow we could  manage to satisfy our ideal boundary conditions and could adjust parameters of the system such that $\tau\sim 1$    than the magnitude of  $ \langle {m}^z_{\rm ind} \rangle$  from eq. (\ref{m_ind}) is determined by parameter $2\pi L_3/e$ such that
\be
\label{estimate}
  \langle {m}^z_{\rm ind} \rangle \sim \frac{2\pi L_3}{e}\sim \frac{L_3 c e}{2\alpha}\sim 1.5\cdot 10^7 \left(\frac{e ~ {\rm cm}^2}{{\rm s}}\right), 
\ee
which is at least  3 orders of magnitude larger than the value (\ref{observed}) of the  magnetic moment generated due to conventional persistent current when the correlation length is determined by the physics of the ring. The crucial  element in our  estimate (\ref{estimate}) is that the key parameter $\tau$ should be order of one, $\tau\sim 1$.  This  
  would guarantee that the vacuum transitions would not be strongly suppressed. We really do not know if it could be   realized  in practice.
  The answer hopefully  could be positive as Aharonov Bohm phase coherence can be maintained at sufficient high temperature,  which can drastically decrease parameter $\tau$ from ({\ref{tau-exp}), see footnote \ref{AB}.
    
 We emphasize once again that  eq.  (\ref{estimate}) describes  a new contribution to the magnetic moment originated from tunnelling transitions between topological $k$ sectors.     It  should be   contrasted with conventional persistent current which also contributes to magnetic moment (\ref{observed}).  These two contributions originated form very different physics:   in case eq. (\ref{observed})  the correlation in the system 
 is achieved  by the dynamics  of the electrons on the boundary, while  in our case it is achieved by the tunnelling transitions between gauge $k$ sectors and described by the vacuum instantons (\ref{replace}) saturating  $\mathcal{Z}_{\rm top} (\tau, \theta_{\rm eff})$. 

\subsection{E\&M Radiation}\label{radiation}
Important comment we would like to make is as follows. Formula (\ref{m_ind}) has been derived assuming that the external field is static. However, formula (\ref{m_ind})  still holds even in the case when  the time dependence is adiabatically slow, i.e.
$(\frac{d B_{\rm ind}}{{dt}})/B_{\rm ind}=\omega $ is much smaller than any relevant   scales of the problem, to be discussed below.
  Therefore, one can use the well -known expressions for the intensity $  \vec{S}$ and total radiated power $I$ for the magnetic dipole radiation when dipole moment (\ref{m_ind}) varies with time:
\be
\label{intensity}
  \vec{S} = I(t) \frac{\sin^2 \theta}{4\pi r^2}\vec{n}, ~~~~~~~~~   I (t) =\frac{2}{3 c^2} \langle \ddot{m}^z_{\rm ind} \rangle^2
\ee
In case when the external magnetic field in the vicinity of $\theta_{\rm eff}\sim 2\pi n$ the behaviour of the induced magnetic moment almost linearly follows  $B_{\rm ext}^z(t) $ as one can see from Figure \ref{fig:bfield}. In particular,  if $B_{\rm ext}^z(t)\sim \cos\omega t$
than $\langle \ddot{m}^z_{\rm ind} \rangle \sim \omega^2\cos \omega t$. In this case one can easily compute the average intensity over large number of complete cycles  with the result 
\be
\label{intensity1}
  \la I \ra \sim \frac{\omega^4}{3 c^2} \langle {m}^z_{\rm ind} \rangle^2,
\ee
where $ \langle {m}^z_{\rm ind} \rangle$ is given by (\ref{m_ind}).

Few comments are in order. First of all, the magnetic dipole radiation can be easily understood in terms of persistent currents \cite{persistent,review_persistent} flowing along the ring. For static external magnetic field the corresponding persistent current $I_0$  is also time independent.  The magnetic dipole moment generated by this current can be estimated as $m_{\rm persistent}\simeq \pi I_0R^2$. When  the external magnetic field starts to fluctuate, the corresponding current $I_0(t)$ as well as magnetic dipole moment $m_{\rm persistent} (t) $ also become time-dependent functions.  It obviously leads to the radiation of real photons which is consistent with our analysis.  However, we should emphasize that the interpretation of this phenomenon (which we coin as  non-stationary TCE) 
  in terms of topological persistent currents (\ref{j}) is the consequential, rather than fundamental explanation. The fundamental explanation, as emphasized in the previous section \ref{interpretation}  is based on topological instanton-like configurations interpolating between $k$ topological sectors. These tunnelling transitions occur in the system even when persistent currents are not generated in the system (for example in absence of external field).

  Our final comment in this subsection is as follows. As we discussed above the  energy for $E\&M$  radiation eventually comes  from  time-dependent external magnetic field. One could suspect that it would be very difficult to discriminate a (non-interesting) direct emission  originated from  $B_{\rm ext}(t)\neq 0$ and the (very interesting) emission resulted from the Maxwell vacuum which itself is excited due to the quantum interference  of the vacuum configurations 
  describing the  topological $| k \ra$ sectors with external magnetic  field.  First (non-interesting) term is represented by $\theta_{\rm eff}=eB_{\rm ext}(t)\pi R^2$ in the parentheses in eq. (\ref{m_ind}), while the second (very interesting) term is represented by term $\sim k$ in eq. (\ref{m_ind}). 
  
  Fortunately, 
  one can easily discriminate between (the very interesting)   emission from the vacuum and  (absolutely non- interesting)  background radiation. 
  The point is that the induced magnetic dipole moment $ \langle {m}^z_{\rm ind}(t) \rangle$ is the periodic function of   $B_{\rm ext}(t)$.
  Exactly at the point      $\theta_{\rm eff}=\pi$    the induced magnetic dipole moment $ \langle {m}^z_{\rm ind} \rangle$ suddenly changes the sign as one can see from  Figure \ref{fig:bfield}. This is a result of complete reconstruction of the ground state 
 in the vicinity of  $\theta_{\rm eff}=\pi$  when  the  level  crossing occurs,  which eventually  results  in the double degeneracy of the system at this point. As this is the key  element of the construction which leads to the important observational consequences   related to the topological features of the system,  we  
  elaborate on this issue with more details  in Appendix \ref{Appendix}.
  
  Therefore, one could slowly change the external field $B_{\rm ext}(t)$
  in vicinity of  $\theta_{\rm eff}=\pi$ which corresponds to the half integer flux, to observe the variation in intensity and  polarization of radiation. The corresponding background radiation must   vary smoothly, while the emission from vacuum should change drastically.  One could hope that these drastic changes may serve as  a smoking gun for discovery of a fundamentally novel type of radiation from topological Maxwell vacuum, similar to DCE.

       \section{Conclusion and Future Directions }\label{conclusion}
         In this work we discussed a number of very unusual features   exhibited by    the   Maxwell theory formulated on  a compact manifold $\mathbb{M}$ with nontrivial topological mapping $\pi_1[U(1)]$, which was coined the  topological vacuum ($\cal{TV}$).  All these  features are originated from the topological portion of the partition function ${\cal{Z}}_{\rm top}(\tau, \theta_{\rm eff})$
  and   can not be formulated in terms of  conventional  $E\&M$  propagating photons  with two physical transverse polarizations. In different words, all effects discussed in this paper have a non-dispersive nature. 
  
             The computations of the present work along with previous calculations of refs. \cite{Cao:2013na,Zhitnitsky:2013hba,Zhitnitsky:2014dra}  imply  that the extra energy (and entropy),   not associated   with any physical propagating degrees of freedom,  may emerge  in the  gauge  systems if some conditions are met. This fundamentally new type  of  energy    emerges as a result of dynamics of pure gauge configurations at arbitrary  large distances. The new idea advocated in this work is that this new type of energy can be, in principle, studied  if one place the system in time-dependent  background. In this case we expect that the vacuum topological configurations can radiate  conventional photons which can be detected and analyzed.   
             
             This unique feature of the system when an extra energy is not related to any physical propagating degrees of freedom was the main  motivation for a proposal   \cite{Zhitnitsky:2013pna,Zhitnitsky:2014aja}  that the  vacuum energy of the Universe may have, in fact,  precisely such non-dispersive  nature\footnote{ This new type  of vacuum energy which can not be expressed in terms of propagating degrees of freedom has been in fact well studied in QCD lattice simulations, see \cite{Zhitnitsky:2013pna} with large number of references on the original lattice results.}. This proposal when an extra energy   can not be associated with any propagating particles  should be contrasted with a conventional description when an extra vacuum energy in the Universe is always associated with some  ad hoc   physical propagating degree of freedom, such as inflaton\footnote{There are two instances in evolution of the universe when the vacuum energy plays a crucial  role.
             First instance   is identified with  the inflationary epoch  when the Hubble constant $H$ was almost constant which corresponds to the de Sitter type behaviour $a(t)\sim \exp(Ht)$ with exponential growth of the size $a(t)$ of the Universe. The  second instance when the vacuum energy plays a dominating role  corresponds to the present epoch when the vacuum energy is identified with the so-called dark energy $\rho_{DE}$
   which constitutes almost $70\%$ of the critical density. In the proposal  \cite{Zhitnitsky:2013pna,Zhitnitsky:2014aja}  the vacuum energy density can be estimated as $\rho_{DE}\sim H\Lambda^3_{QCD}\sim (10^{-4}{\rm  eV})^4$, which is amazingly  close to the observed value. }.

    Essentially, the  proposal   \cite{Zhitnitsky:2013pna,Zhitnitsky:2014aja}  identifies the observed vacuum  energy with the Casimir type energy, which however is originated not from dynamics of the physical   propagating degrees of freedom, but rather, from the dynamics of the topological sectors  which  are always present in gauge systems, and which are highly sensitive to arbitrary large distances. Furthermore, the radiation from the vacuum in a time-dependent background (which is  the main subject of this work) is very similar  in all respects to the radiation which might be  responsible for  the  end of inflation in that proposal, see  \cite{Zhitnitsky:2014aja} for the details. The present study, in fact,  is   motivated by  the cosmological ideas     \cite{Zhitnitsky:2014aja} which hopefully can be    tested  in a tabletop experiment when the vacuum energy in time-dependent background can be transferred to real propagating degrees of freedom as suggested in section \ref{radiation}. In cosmology the corresponding period plays a crucial role and calls the reheating epoch which follows  the inflation when the vacuum energy is the dominating component of the Universe. 
  
   To conclude, the main point of the present studies is that the radiation may be generated from the  vacuum configurations describing the tunnelling transitions between $|k\ra$ sectors, rather than from physical propagating degrees of freedom, 
   which  would correspond to conventional DCE when the virtual photons become real photons in a time dependent background.
    This is precisely the difference between DCE and non-stationary TCE considered in the present work.

 \section*{Acknowledgements} 
 I am thankful to Maxuim Chernodub for discussions and useful comments.
   I am also thankful to Alexei  Kitaev for long and interesting discussions on relation of the persistent currents \cite{persistent, review_persistent} and the topological configurations saturating  $\mathcal{Z}_{\rm top}$.
    This research was supported in part by the Natural Sciences and Engineering Research Council of Canada.


\appendix

\section{Classification of the vacuum states, Degeneracy, and the Topological order.}\label{Appendix}
The main goal of this Appendix is to review and elaborate on  important property of degeneracy of the system under study. As   suggested  in section \ref{radiation} 
 the feature of degeneracy may  play an important  role in  discrimination of novel and  interesting effect of emission of real photons from vacuum (as a result of  the non-static  Topological Casimir Effect) from the background radiation.

The starting point is to analyze  the symmetry properties of  partition function  ${\cal{Z}}_{\rm top}(\tau, \theta_{\rm eff})$ defined by eq.(\ref{replace}). 
 One can easily observe that  the $\theta_{\rm eff}=\pi$ is very special point as even and odd $``n"$- terms in the dual representation for ${\cal{Z}}_{\rm top}(\tau, \theta_{\rm eff})$  contribute equally to the partition function. It obviously   leads to the degeneracy of the system at  $\theta_{\rm eff}=\pi$.
 The conventional, non-topological,    part of the   partition function ${\cal{Z}}_{\rm quant}(\tau, \theta_{\rm eff})$ is not sensitive 
 to the topological sectors at all, as discussed in section \ref{construction}. Therefore,  this property of degeneracy is an exact feature  of the system. This double degeneracy implies that ${\mathbb{Z}_2} $ symmetry is  spontaneously  broken. 
 
 What is the symmetry which  is spontaneously broken? What is order parameter which classifies two  physically distinct  states? One can not formulate the corresponding symmetry breaking effect  in terms of any local operators and their vacuum expectation values  as argued  in \cite{Zhitnitsky:2013hba}.  Rather, a proper  classification of the ground state is formulated in terms of non-local operators. 
 Indeed, a corresponding  order parameter which characterizes the system is 
   \be
\label{theta_pi}
 &&  \langle \frac{e}{2\pi}\oint A_i dx_i\rangle_{\theta_{\rm eff}=\pi-\epsilon}  =+\frac{1}{2}\\
  &&   \langle \frac{e}{2\pi}\oint A_i dx_i\rangle_{\theta_{\rm eff}=\pi+\epsilon}=-\frac{1}2  \nonumber
\ee
where computations should be carefully  carried out by approaching $\theta_{\rm eff}=\pi$ in partition function ${\cal{Z}}_{\rm top}(\tau, \theta_{\rm eff})$ 
 from two opposite  sides of the $\theta_{\rm eff}=\pi\pm \epsilon$ as discussed in \cite{Zhitnitsky:2013hba}. 
  This classification is very different from   the conventional Landau classification based on broken symmetries when 
  a  system is characterized by  some   expectation value of a local operator. One should note that 
  a  similar classification is known to emerge  in topologically order systems, e.g. topological  insulators at $\theta=\pi$, see recent reviews \cite{Cho:2010rk,Wen:2012hm,Sachdev:2012dq,Cortijo:2011aa,Volovik:2011kg}. In fact the classification  (\ref{theta_pi})   is a very particular example of a much more generic framework recently discussed in    \cite{Gaiotto:2014kfa} when the classification of a  ground state  is based on some  global rather than   local observables. 
  
  We want to emphasize  that classification (\ref{theta_pi}) is not a unique feature of the abelian gauge theory. Similar classification of the ground states also emerges in non-abelian gauge field theories. We would like to mention in this Appendix  just one specific example, the so-called ``deformed QCD" model \cite{Unsal:2008ch}  because its close relation to real QCD, and most importantly, because this model shows a number of  features relevant for the present studies, such as the generating  of the vacuum energy which is not associated with any propagating degrees of freedom (the Topological  Casimir Effect). Furthermore, it may also exhibit non-static TCE discussed in section \ref{radiation} when the radiation of real particles occurs as a result of a time dependent background, which could have profound consequences for cosmology, as we already mentioned in concluding section \ref{conclusion}.

 The ``deformed QCD"  model is  a weakly coupled gauge theory, when the analytical computations can be performed in theoretically  controllable  way.  Though it is a weakly coupled gauge theory,  it nevertheless preserves all the crucial elements of strongly interacting QCD, including confinement, nontrivial $\theta$ dependence, degeneracy of the topological sectors, etc.  Furthermore, it has been claimed \cite{Unsal:2008ch} that there is no any phase transition in passage from weakly coupled deformed QCD to strongly coupled QCD,    such that  this model allows to address and answer a number of highly non-trivial questions. 
  Important for the present work topic   which will be reviewed  here is the classification of the ground states in this system.

As described in \cite{Unsal:2008ch} the proper infrared description of the theory is a dilute gas of $N$ types of monopoles characterized by their   fractional topological charge $1/N$ and  magnetic charges, which are proportional to the simple roots and affine root $\alpha_{a} \in \Delta_{\mathrm{aff}}$ of the Lie algebra of  the $SU(N)$  gauge group. 
One can explicitly construct  \cite{Thomas:2011ee}  the creation operator  
${\cal{M}}_a (\mathbf{x})=e^{i \alpha_{a} \cdot \bm{\sigma} (\mathbf{x})}$
 for a monopole of type $a$ at point $\mathbf{x}$, and compute its expectation value with the following result    \cite{Bhoonah:2014gpa}: 
\be
\label{magnetization}
\langle{\cal{M}}_a (\mathbf{x})\rangle_m =\exp{\left[{ i\frac{\theta+2\pi m }{N}}\right]},
\ee 
where $m$ is integer which classifies a specific  ground state. These distinct vacuum states classified by $m$ are not degenerate as it normally happens in supersymmetric field theories. Nevertheless, some  degree of  degeneracy (corresponding to integers $m$ and $-m$ for $\theta=0$) still holds.   However, these states, being the local minima of the system  are metastable states, rather than absolutely stable states,  because they can decay  to the ground state $m=0$ through the tunnelling transition. The corresponding decay rate can be expressed in terms of the  of the domain walls which separate these vacua  \cite{Bhoonah:2014gpa}. There is a large number of different types of domain walls in the system.  Their classification   is uniquely fixed  by the classification  of corresponding vacua (\ref{magnetization}).  
  
 It is also instructive to observe how the confinement in the vacuum states  (\ref{magnetization}) is realized.  It turns out that the confinement 
 in this system is  a result of the condensation of the fractionally charged monopoles, as usual for $\theta=0, m=0$. 
 However,  the  corresponding vacuum structure  for $m\neq 0$ is less trivial as it represents a specific superposition of $N$ different monopole condensates. To be more specific: the condensation of the  $N$ different monopole types is described by the same $N$ component $ \bm{\sigma} (\mathbf{x})$ field  which describes the confinement  for  $m=0$ state. 
However, different monopole  condensates   form  a coherent  superposition   \cite{Bhoonah:2014gpa}  shifted by a phase such that the corresponding   magnetization $\langle{\cal{M}}_a (\mathbf{x})\rangle_m$ receives a non-trivial phase (\ref{magnetization}) for each specific superposition.   In other words,  different vacuum states classified by parameter $m$ from (\ref{magnetization}) correspond to the different superpositions of $N$ different monopole condensates \footnote{It is often (wrongly) interpreted that the confinement for  metastable vacuum states  is due to the condensation of the dyons carrying the electric charges along with the magnetic charges.  This (incorrect) interpretation based on observation that the topological density operator is the product of electric and magnetic fields, $\vec{E}\cdot\vec{B}$. One should remember, however, that  the monopoles in this construction are pseudoparticles living in 4d Euclidean space-time, rather than static 3d objects. The finite action   and  finite topological charge $1/N$ for these objects is  a result of wrapping of the monopole's path along the Euclidean time direction $ {\mathbb{S}}^1$ with nontrivial holonomy. 
These objects do not carry a conventional static electric charge;  nevertheless, they do carry the topological charges defined in 4d Euclidean space-time. 

There is no fundamental  difference between the vacuum structures  for the metastable states and the lowest energy  state. In all cases the confinement is realized as a condensation  of the same monopoles, not dyons (by the reasons mentioned above it is more appropriate to use term ``percolation"  rather than ``condensation" which is normally used for truly static 3d objects in condensed matter physics). The only difference in  structure between the metastable and the ground states is that  $N$ different  monopole's condensates describing  a metastable vacuum state is represented by a superposition which includes   a relative phase shift  between different condensates, in contrast  with the superposition  which describes  the ground vacuum state when this phase identically vanishes, see \cite{Bhoonah:2014gpa} for the details.}, while the effective long range field describing the dynamics of these monopolies, the $ \bm{\sigma} (\mathbf{x})$-field, remains always the same for all $m$. This picture resembles in many ways the picture of the so-called oblique confinement suggested long ago by 't Hooft \cite{thooft} for some special  values of the $\theta_m=\frac{2\pi m}{N}$.

Our final remark of this Appendix is that  the Maxwell gauge system defined on a compact manifold and studied in the present  work, as well as non-abelian gauge ``deformed QCD" model reviewed in this Appendix, show a number of other features (along with property of degeneracy already mentioned here)  which are inherent characteristics   of the  topologically ordered phases. We refer to   \cite{Zhitnitsky:2013hba,Zhitnitsky:2014dra} for corresponding arguments related   to  the $U(1)$ Maxwell gauge system formulated on a compact manifold, and to \cite{Zhitnitsky:2013hs} with corresponding arguments applied to  the $SU(N)$ ``deformed QCD" model  for the details.


\begin{thebibliography}{99}
    
\bibitem{Cao:2013na} 
  C.~Cao, M.~van Caspel and A.~R.~Zhitnitsky,
  Phys.\ Rev.\ D {\bf 87}, 105012 (2013)
  [arXiv:1301.1706 [hep-th]].
 
\bibitem{Zhitnitsky:2013hba} 
  A.~R.~Zhitnitsky,
  Phys.\  Rev.\  D {\bf 88}, { 105029} (2013)
   [arXiv:1308.1960 [hep-th]].
   
\bibitem{Zhitnitsky:2014dra} 
  A.~Zhitnitsky,
  Phys.\ Rev.\ D {\bf 90},  105007 (2014)
  [arXiv:1407.3804 [hep-th]].

  \bibitem{Casimir}
   H. B. G. Casimir, Kon. Ned. Akad. Wetensch. Proc. {\bf 51}, 793 (1948).
  
\bibitem{Cho:2010rk} 
  G.~Y.~Cho and J.~E.~Moore,
  Annals Phys.\  {\bf 326}, 1515 (2011)
  [arXiv:1011.3485 [cond-mat.str-el]].

\bibitem{Wen:2012hm} 
  X.~-G.~Wen,
  arXiv:1210.1281 [cond-mat.str-el].
  
\bibitem{Sachdev:2012dq} 
  S.~Sachdev,
  arXiv:1203.4565 [hep-th].
  
   
\bibitem{Cortijo:2011aa} 
  A.~Cortijo, F.~Guinea and M.~A.~H.~Vozmediano,
  J.\ Phys.\ A {\bf 45}, 383001 (2012)
  [arXiv:1112.2054 [cond-mat.mes-hall]].

\bibitem{Volovik:2011kg} 
  G.~E.~Volovik,
  Lecture Notes in Physics, {\bf 870}, 343 (2013)
  [arXiv:1111.4627 [hep-ph]].
  
  \bibitem{DCE}
  G. T. Moore, J. Math. Phys. {\bf 11}, 2679 (1970); \\
  S. A. Fulling and P. C. W. Davies, Proc. R. Soc. Lond. {\bf A 348}, 393 (1976); \\
  P. C. Davies and S. A. Fulling, Proc. Roy. Soc. Lond. {\bf A 356}, 237 (1977).
  
  \bibitem{DCE-review}
  G. Barton and C. Eberlein, Ann. Phys. {\bf 227}, 222 (1993);\\
  M. Kardar et al., Rev. Mod. Phys. {\bf 71}, 1233 (1999);\\
       V. V. Dodonov, pp. 309 in Modern Nonlinear Optics, Part 3, ed. M. W. Evans, Adv. Chem. Phys. Series, Vol. {\bf 119} (Wiley, New York, 2001). 
  
  
  \bibitem{DCE-exp}
  C. M. Wilson, G. Johansson, A. Pourkabirian, M. Simoen, J. R. Johansson, T. Duty, F. Nori
and P. Delsing, Nature, {\bf 479}, 376 (2011);\\
  P.Lahteenmaki, G.S.Paraoanu, J.Hassel, and P.J.Hakonen,
Proc. Natl. Acad. Sci. U.S.A. {\bf 110}, 4234 (2013).

\bibitem{persistent}
M. Buttiker, Y. Imry, and R. Landauer, Phys. Lett.  A {\bf 96},
365 (1983);\\
I. O. Kulik, Pis'ma Zh. Eksp. Teor. Fiz. {\bf 11}, 407 (1970); JETP
Lett. {\bf 11}, 275 (1970).


\bibitem{review_persistent}
 Y. Imry, Introduction to Mesoscopic Physics, 2nd ed. (Oxford University Press, New York, 2008);\\
 E. Akkermans and G. Montambaux, Mesoscopic Physics of
Electrons and Photons, reissue ed. (Cambridge University Press, 2011);\\
William Ennis Shanks,  [arXiv:1112.3395 [cond-matt.mes-hall]].

\bibitem{persistent-exp}
V. Chandrasekhar, R. A. Webb, M. J. Brady, M. B. Ketchen, W. J. Gallagher, and A. Kleinsasser, 
Phys. \ Rev.  \ Lett.,  {\bf 67},   3578 (1991).

\bibitem{persistent-temp}
M. Tsubota, K. Inagaki, and S. Tanda
EPL {\bf 97}, 57011 (2012)[arxiv:0906.5206 [cond-matt.mes-hall]].



\exclude{ 
\bibitem{Jaffe:2005vp} 
  R.L.~Jaffe,
  Phys.\ Rev.\ D {\bf 72}, 021301 (2005)
  [hep-th/0503158].

 }
    
 
   
\bibitem{Zhitnitsky:2013pna} 
  A.~R.~Zhitnitsky,
  Phys.\ Rev.\ D {\bf 89}, 063529 (2014)
  [arXiv:1310.2258 [hep-th]].
  
\bibitem{Zhitnitsky:2014aja} 
  A.~R.~Zhitnitsky,
  Phys.\ Rev.\ D {\bf 90}, 043504 (2014)
  [arXiv:1404.5965 [hep-ph]].
  
  \exclude{
\bibitem{Zhitnitsky:2014ria} 
  A.~R.~Zhitnitsky,
  ``The topological long range order in QCD. Applications to heavy ion collisions and cosmology,'' 
  arXiv:1411.2606 [hep-ph].
 } 
\bibitem{Gaiotto:2014kfa} 
  D.~Gaiotto, A.~Kapustin, N.~Seiberg and B.~Willett,
  ``Generalized Global Symmetries,''
  arXiv:1412.5148 [hep-th].
  
\bibitem{Unsal:2008ch} 
  M.~Unsal and L.~G.~Yaffe,
  Phys.\ Rev.\ D {\bf 78}, 065035 (2008)
  [arXiv:0803.0344 [hep-th]].
  
\bibitem{Thomas:2011ee} 
  E.~Thomas and A.~R.~Zhitnitsky,
  Phys.\ Rev.\ D {\bf 85}, 044039 (2012)
  [arXiv:1109.2608 [hep-th]].
  
\bibitem{Bhoonah:2014gpa} 
  A.~Bhoonah, E.~Thomas and A.~R.~Zhitnitsky,
  Nucl.\ Phys. \ B {\bf 890}, 30 (2015)
  [arXiv:1407.5121 [hep-ph]].
  
  \bibitem{thooft}
  G. 't Hooft, Nucl.\ Phys.\  B {\bf 153} , 141 (1979),\\
  G. 't Hooft, Nucl. \ Phys. \ B {\bf 190}, 455 (1981).
  
\bibitem{Zhitnitsky:2013hs} 
  A.~R.~Zhitnitsky,
  Annals Phys.\  {\bf 336}, 462 (2013)
  [arXiv:1301.7072 [hep-ph]].
  
    \end{thebibliography}
\end{document}